\documentclass[11pt]{cernrep}
\usepackage{graphicx,cancel,axodraw}

\def\beq{\begin{equation}}
\def\eeq{\end{equation}}
\def\beeq{\begin{eqnarray}}
\def\beeqn{\begin{eqnarray*}}
\def\eeeq{\end{eqnarray}}
\def\eeeqn{\end{eqnarray*}}

\def\e{\varepsilon}

\newcommand{\no}{\nonumber}

\def\frac#1#2{ {{#1} \over {#2} }}

\def\p{\partial}

\begin{document}
\title{SYM $\cal{N}$=4 in light-cone gauge and the ``bridge'' identities}
\author{A. Bassetto and G. De Pol}
\institute{Dipartimento di Fisica ``G.Galilei", Via Marzolo 8, 35131
Padova, Italy\\
INFN, Sezione di Padova, Italy\\ \texttt{(bassetto, depol@pd.infn.it)} }

\maketitle
\begin{abstract}
The light-cone gauge allows to single out a set of ``transverse'' fields (TF),
whose Green functions are free from UV divergences in SYM $\cal{N}$=4.
Green functions with external lines involving the remaining fields 
do instead exhibit divergences: indeed
those fields can be expressed, by solving their equations of motion, as
composite operators in terms of ``transverse'' fields. A set of exact identities
(bridge identities) automatically realize their insertions in a path-integral 
formulation.
\end{abstract}

\vskip 0.5truecm

\noindent
DFPD/TH 07-07



\section{Introduction}

\noindent
Long ago it has been proved that the maximally extended supersymmetric Yang-Mills
theory SYM $\cal{N}$=4 \cite{bss,gso}, when quantized in light-cone gauge, possesses 
a set of Green
functions, the ``transverse sector'' (TS) \cite{lbnl}, which are (perturbatively) 
UV-finite at any order in the loop expansion \cite{mand,bln} \footnote{To reach this goal,
as firstly pointed out by Mandelstam \cite{mand}, a correct
definition of the vector propagator is essential.}, a result which is enough to conclude 
that the $\beta$-function vanishes. In turn this property,
being gauge invariant, entails
deep consequences on important features, like for instance anomalous dimensions 
and ADS/CFT duality \cite{lipa}. Still the theory exhibits divergences in the set of 
the remaining Green functions (the non-transverse sector) (NTS). 

Quite generally, the properties of an arbitrary 
Yang-Mills theory (YMT), once correctly 
quantized in the light-cone gauge \cite{basla}, have been extensively worked out and 
renormalization has been proved
recursively to any order in the loop expansion, providing a full control
of one-particle irreducible vertices and of Green functions  \cite{basdal,libro}.
 
The way in which divergences are produced in SYM $\cal N$=4 and how they 
can be handled
seem not to have been exhaustively examined so far. This may be relevant also 
in connection with some persisting uncertainties in the regularization procedure 
when supersymmetry Slavnov-Taylor identities are considered \cite{avd,ted}. 

The main result of this paper is to generalize a set of identities (BI), 
sketched for the pure YMT in ref.\cite{diff} and 
introduced for QCD in ref.\cite{germ}, to the supersymmetric case $\cal{N}$=4. 
They allow to express the Green functions of the NTS in terms of 
integrals involving transverse Green functions; in turn the latter can be 
computed by means for instance of supergraphs and do not require a regularization 
procedure, just being UV-finite \cite{bln}\footnote{Divergences occur in the (TS) when
composite operators are considered.}.

These identities hold for exact amplitudes, {\it i. e.} are not confined to
perturbation theory. In such a context they hold order by order in the coupling 
constant expansion.

\vskip .5 truecm

Regularization and renormalization of an arbitrary YMT in light-cone gauge are
briefly recalled in Sect.2; in so doing the relevant notations are introduced.
The lagrangian density of SYM $\cal{N}$=4 is given explicitly, together
with the projection operators singling out the TF and the NTF.
General renormalization results are particularized to this case
and the argument for the finiteness of the Green functions in the TS is briefly recalled
\cite{bdal}.
In Sect.3 the BI are derived and their mathematical structure is discussed,
in particular their involving peculiar composite operators which are eventually 
responsible
of the appearance of divergences. In Sect.4 the simplest BI are perturbatively
checked at the one-loop level, in order to show how they operate in actual calculations. 
Sect.5 contains our final remarks and conclusions, while figures, figure captions
and technical details are given in the Appendix. 

\section{The renormalization in light-cone gauge}

\noindent
The quantization and the renormalization of a generic YMT  in light-cone
gauge have been extensively studied in the eighties \cite{libro,bec}. After using the
Mandelstam-Leibbrandt 
prescription in the vector propagator 
\beeq
\label{manlei}
&&\tilde G_{\mu\nu}^{(0),ab}(p)= \frac{-i\delta^{ab}}{p^2+i\epsilon}
\Big[g_{\mu\nu}-\frac{n_{\mu} 
p_{\nu}+n_{\nu}p_{\mu}}{n\cdot p}\Big], \qquad\qquad  n_{\mu}=\frac{1}{\sqrt{2}}
(1,0,0,1), \\ \nonumber 
&&\frac{1}{n\cdot p}\equiv \frac{\hat n\cdot p}{\hat n\cdot p \, n\cdot p+i\e}=
\frac{1}{n\cdot p+i\e \mbox{sign}(\hat n\cdot p)}, \qquad \hat n_{\mu}=\frac{1}{\sqrt{2}}
(1,0,0,-1)
\eeeq
proposed in \cite{mand,lei},
and derived from a canonical quantization in \cite{basla},
the Wick rotation is possible in Feynman diagram calculations and thereby
the usual power counting criterion for convergence. The novelty concerns the
appearance of non-polynomial singularities in the residues of the poles 
when using dimensional regularization \footnote{Actually, to preserve supersymmetry,
we rather use dimensional reduction \cite{avd,ted}.};
their origin can be traced back to the gauge (``spurious'') singularity
of the vector propagator. 

In spite of the fact that non-polynomialities threaten dimensional
counting, it has been proved that all non-polynomial counterterms are
encoded, at any order in the loop expansion, in a unique non-local structure which
provides the required subtractions \cite{basdal}. Its presence affects the
proper vertices, but non-localities cancel in the Green functions and
{\it a fortiori} in S-matrix elements where gauge and Lorentz invariance
are correctly recovered. These renormalization results have been successfully
tested in quite non-trivial two-loop calculations concerning Wilson loops
and partonic anomalous dimensions \cite{korc,vog}.

The theory is even simpler when the maximally extended supersymmetry $\cal{N}$=4
is introduced, although the S-matrix cannot be defined in this case. 
To be concrete, let us
write explicitly its lagrangian density \cite{gso}        
\beeq \label{lagra}
{\cal L}&=&-\frac{1}{4}F_{\mu \nu}^{a}F^{\mu \nu ,a}+ \frac{1}{2}
(D_{\mu} 
\phi_{r}^{a})^2 + \frac{1}{2}(D_{\mu} \chi_{r}^{a})^2 +\nonumber \\
&-&\frac{g^2}{4}
\left[(f^{abc} \phi_{r}^{b} \phi_{t}^{c})^2+(f^{abc} \chi_{r}^{b} 
\chi_{t}^{c})^2 +2 (f^{abc} \phi_{r}^{b} \chi_{t}^{c})^2\right]+\nonumber\\
&+&\frac{i}{2}\bar{\lambda}_{m}^{a}(\cancel{D} \lambda_{m})^a -\frac{1}{2}g f^{abc}
\bar{\lambda}_{m}^{a} (\alpha_{mn}^{r} \phi_{r}^{b}+ \gamma_5 \beta_{mn}^{r}
 \chi_{r}^{b})\lambda_{n}^{c} -\Lambda^{a} n_{\mu}A^{\mu,a}
\eeeq
where, as usual,
\beq
\label{campo}
F_{\mu\nu}^{a}=\p_{\mu}A_{\nu}^{a}-\p_{\nu}A_{\mu}^{a}+g f^{abc}A_{\mu}^{b}A_{\nu}^{c}
\eeq
and $A_{\mu}^{a},\lambda_m^a,\phi_r^a,\chi_r^a$ represent respectively vector potentials,
Majorana spinors, scalar and pseudoscalar fields, all lying in the adjoint representation
of the gauge group. $\Lambda^a$ are Lagrange multipliers and $n_{\mu}$ is the 
gauge-fixing vector. ``Color'' and internal indices will be sometimes understood.

The matrices $\alpha_{mn}^r$ and $\beta_{mn}^r$, $m,n=1,...,4, \,\,\, r=1,2,3$ are
\beeq 
\label{glio}
\begin{array}{ccc}
\alpha^{1}=\left(\begin{array}{rr}
0 & \sigma^{1}\\
-\sigma^{1} & 0\\
\end{array}\right) &
\alpha^{2}=\left(\begin{array}{rr}
0 & -\sigma^{3} \\
\sigma^{3} & 0\\
\end{array}\right) &
\alpha^{3}=\left(\begin{array}{rr}
i \sigma^{2} & 0\\
0 & i \sigma_{2}\\
\end{array}\right)
\\
\beta^{1}=\left(\begin{array}{rr}
0 & i \sigma^{2}\\
i \sigma^{2} & 0\\
\end{array}\right) &
\beta^{2}=\left(\begin{array}{rr}
0 & 1 \\
-1 & 0\\
\end{array}\right) &
\beta^{3}=\left(\begin{array}{rr}
-i \sigma^{2} & 0\\
0 & i \sigma_{2}\\
\end{array}\right)\\
\end{array}
\eeeq
and satisfy \footnote{When performing a dimensional reduction this algebra
has to be suitably deformed \cite{avd}: $\alpha^r\alpha^r=\beta^r\beta^r=\omega-5$.}: 
\beeq
\label{gliozzialgebra}
\{\alpha_{i},\alpha_{j}\}&=&\{\beta_{i},\beta_{j}\}=-2 \delta_{i,j}\nonumber \\
\left[ \alpha_{i},\beta_{j}\right]&=& 0,\qquad \alpha^r\alpha^r=\beta^r\beta^r=-3.
\eeeq
In light-cone gauge some components (NTF) turn out to be gaussian and can be expressed
in terms of a subset of independent fields, namely \cite{mand}
\beeq
\label{man}
A_{\alpha}^{a}\,,\,\lambda_{+,m}^{a}\equiv P_{+}\lambda_{m}^{a}
\,,\,\phi_{r}^{a}\,,\,\chi_{r}^{a}
\eeeq
where the projection operators $P_{\pm}$ are defined as
\beq\label{proj}
P_{\pm}=\frac{\gamma^{\mp}\gamma^{\pm}}{2},\qquad \gamma^{\pm}\equiv\frac{\gamma^0 
\pm\gamma^3}{\sqrt{2}},
\eeq
and Lorentz indices from the beginning of the Greek alphabet are taken to be transverse 
$\alpha\,,\,\beta=1\,,\,2\,.$
Sometimes those fields are referred to as ``physical'' ones; we have decided to call 
them ``transverse'' (as a matter of fact the fermionic decomposition $\lambda=
\lambda_+ + \lambda_-$ is obviously gauge dependent and the $\lambda_-$ are certainly
not less physical than the $\lambda_+$).

The TF can be recast in a single chiral-type multiplet \cite{bln}; then, using superfield
Feynman diagrams, it has been shown that all Green functions turn out to be
UV convergent \cite{mand}, since the Mandelstam-Leibbrandt 
prescription in the vector propagator 
guarantees the validity of the power counting, which in turn is an essential 
ingredient of the proof of finiteness.

Let us now rephrase the analysis of \cite{basdal} 
where it has been proved that the 
most general counterterms required in the total Lagrangian can be reproduced by
means of the redefined fields and charge

\beeq \label{bare}
A_{\mu}^{(0)} &=& Z_3^{1/2}[A_{\mu}-(1-\tilde Z_3^{-1})n_{\mu}\Omega], \no \\
\lambda^{(0)} &=& Z_2^{1/2}(P_+ +\tilde Z_2 P_-)\lambda,\no \\
\phi_r^{(0)}  &=& Z_{\phi}^{1/2}\phi_r, \no  \\
\chi_r^{(0)}  &=& Z_{\chi}^{1/2}\chi_r, \no \\
\Lambda^{(0)} &=& Z_3^{-1/2}\Lambda, \no \\
g_0  &=& Z_3^{-1/2}g, 
\eeeq

where 

\beq \label{nonl} 
n_{\mu}D^{\mu}\Omega \equiv n_{\mu}\hat n_{\nu}F^{\mu\nu},
\eeq
$D^{\mu}$ being the covariant derivative in the adjoint representation.
In the present case $Z_3=1$ (in a minimal subtraction scheme) and $Z_2,
Z_{\phi},Z_{\chi}$ are also one, thanks to the supersymmetry of the TS,
at any order in the loop expansion.
The two remaining renormalization constants $\tilde Z_3$ and  $\tilde Z_2$
do instead diverge, in spite of the vanishing of the $\beta$-function:
\beq \label{renco}
\tilde Z_2 = 1-g^2 C_{Ad}\frac{1}{8\pi^2(2-\omega)}+O(g^4),\,\,\,
\tilde Z_3 = 1+g^2 C_{Ad}\frac{1}{8\pi^2(2-\omega)}+O(g^4),
\eeq
$C_{Ad}$ being the quadratic Casimir operator in
the adjoint representation of the ``color'' group.

The results of one-loop calculation of the divergent parts of self-energy
diagrams are presented in \cite{bdal}. In Sect.4 and in the Appendix we
generalize those results including all finite contributions as well. To our
knowledge it is the first time they appear in the literature.
Then we  verify that the transverse components (TC) of the Green functions
are indeed finite. The non-transverse components (NTC) do instead diverge.

In QCD the divergent part $\hat \Gamma$ of the regularized effective action
contains the gauge dependent and Lorentz non covariant expression:
\beq \label{div}
\hat \Gamma_{nc}=  i\tilde{a}_{2} \bar{\psi}\left(
\frac{n \cdot D \cancel{\hat{n}}-\hat{n}\cdot D \cancel{n}}{n\cdot \hat{n}}
\right)\psi+
\tilde{a}_{3}\frac{\Omega^{a}n^{\mu}}{n \cdot \hat{n}}\left[
\left(D^{\nu}F_{\mu \nu}\right)^{a} - g \mu^{2-\omega} \left(
\bar{\psi}\tau^a\gamma_{\mu} \psi \right)\right],
\eeq
where $\tilde a_2$ and $\tilde a_3$ are divergent coefficients when the number of 
space-time dimensions $2\omega$ tends to 4.
It is easy to realize that the Lorentz non-covariant parts do not contribute 
to S-matrix elements as those extra pieces are proportional to the classical 
equations of motions \cite{vor}\footnote{The coefficient of
$\tilde a_2$ can be shown to vanish taking the Dirac equation for $\lambda$ and 
$\bar\lambda$ into account; the coefficient of $\tilde a_3$ contains in square
brackets the equation of motion for $F_{\mu\nu}$.}.

In the present case, where the S-matrix cannot be consistently defined,
one should resort to different ``observable quantities'', like, for instance,
Wilson loops. It is expected that singular Lorentz non covariant contributions
cancel here as well, but this should be proved. We plan to investigate this 
interesting issue shortly.

\section{The bridge identities}

\noindent
Now we want to exploit the consequences on the Green functions generated
by the presence in the Lagrangian of field components which appear only
quadratically and with the coefficient of the quadratic term being 
{\it purely kinematical}, namely not involving other fields.

After adding to the original Lagrangian a term containing sources
\beq\label{sour} 
{\cal L}_{\mbox{s}}=\frac{1}{2}(\bar{\lambda}_{m}^{a} \zeta_{m}^{a} +\bar{\zeta}_{m}^{a}
\lambda_{m}^{a})+J^{\mu\,a}A_{\mu}^{a}
+h_{r}^{a} \phi_{r}^{a} +k_{r}^{a} \chi_{r}^{a}
\eeq
the generating functional W becomes
\beq\label{integ}
W=
\int {\cal D}[A_{\mu}^{a}\,,\,\lambda_{m}^a\,,\, \phi_{r}^{a}\,,\,
\chi_{r}^{a}] \exp i \int ({\cal L}+{\cal L}_{\mbox{s}}) d^4 x . 
\eeq
Then the gaussian components can be explicitly integrated over leading to
\beeq\label{twoco}
W&=&\exp \left[\frac{i}{2}\int (J^+ \p_{-}^{-2}J^+ + \bar{\zeta}_{+}
\frac{i \gamma^{+}\p_{+}}{\p_{+}\p{-}}\zeta_{+})d^4 x\right] \nonumber\\
&&\int {\cal D}[A_{\alpha}, \lambda_{+}] \exp i \int ({\cal L}_{\mbox{eff}}+
{\cal L}_{\mbox{mix}}+{\cal L}_{\mbox{tr}})d^4 x
\eeeq
where
\beeq
{\cal L}_{\mbox{eff}}&=&-\frac{1}{4} F_{\alpha \beta}^{a}F_{\alpha \beta}^{a}
+\p_{+}A_{\alpha}\p_{-}A_{\alpha} -\frac{1}{2}K^2 +\frac{i}{4}
\bar{\Theta}\p_{-}^{-1}\gamma^{+}\Theta ,\\
{\cal L}_{\mbox{mix}}&=& J^+ \p_{-}^{-1}K +\frac{i}{4}
(\bar{\zeta}_{+}\p_{-}^{-1}\gamma^{+}\Theta + \bar{\Theta} \p_{-}^{-1}
\gamma^{+} \zeta_{+}), \\
{\cal L}_{\mbox{tr}}&=&J^{\alpha}A_{\alpha}+\frac{1}{2}(\bar{\zeta}_{-}
\lambda_{+}+\bar{\lambda}_{+}\zeta_{-})+(h \phi +k \chi)\,, 
\eeeq
and
\beeq
K^{a}&=&\p_{-}^{-1}\left[(D_i \p_{-}A_{i})^{a}-\frac{i g}{2}f^{abc}
\bar{\lambda}_{+,m}^{b}\gamma^{+}\lambda_{+,m}^c+\right.\nonumber\\
&+&\frac{g}{2}f^{abc}(\phi^{b}_{r}
\p_{-}\phi^{c}_{r}-\p_{-}\phi^{b}_{r}\phi^{c}_{r})+
 \left.+\frac{g}{2}f^{abc}(\chi^{b}_{r}
\p_{-}\chi^{c}_{r}-\p_{-}\chi^{b}_{r}\chi^{c}_{r})\right],\\
\Theta_{m}^{a}&=& 
i \gamma^{\alpha}(D_{\alpha}\lambda_{+,m})^a -g f^{abc}
(\alpha_{mn}^{r} \phi_{r}^{b}+ \gamma_5 \beta_{mn}^{r}
 \chi_{r}^{b})\lambda_{+,n}^{c}=\\
&=&(i \cancel{\p}_{\perp}\lambda_{+,m}^{a}
+i g f^{abc}\gamma^{\alpha}A_{\alpha}^{b}\lambda_{+,m}^{c}) -g f^{abc}
(\alpha_{mn}^{r} \phi_{r}^{b}+ \gamma_5 \beta_{mn}^{r}
 \chi_{r}^{b})\lambda_{+,n}^{c}\,.\nonumber
\eeeq
Eqs.(\ref{integ}) and (\ref{twoco}) provide two equivalent ways of
formulating the theory (the so-called four- and two- component
formulations \cite{cap}). In a perturbative context they give rise 
to different Feynman rules
and to graphs with different topologies. In particular in the two-component 
formulation propagators and vertices exhibit peculiar
non-polynomialities which prevent a direct proof of renormalization. 
The rules in the four-component formulation are instead 
the standard ones in light-cone gauge.
Renormalization can here be proved following general theorems 
and then translated in the two-component formulation \cite{germ}.

From a computational point of view the four-component formulation
has proved in the past to be by far simpler. On the other hand,
in SYM $\cal{N}$=4, calculations in the TS can be performed by
supergraph techniques, {\it i.e.} considering only transverse fields.

Functional differentiation with respect to sources multiplying
NTF obviously generate Green functions with non-transverse
external lines. The crucial remark is now that a differentiation with
respect to a non-transverse source can be traded, thanks to eq.(\ref{twoco}),
with differentiations with respect to transverse sources, giving rise
to bridge identities (BI) between two- and four-component formulations. 
Those identities, proposed 
for QCD in \cite{germ},  can be straightforwardly generalized to the
SYM $\cal{N}$=4 case at hand. They allow to express Green functions with some
non-transverse external lines in terms of purely transverse Green functions.
In this context, where the TS is UV-finite and therefore 
needs not be renormalized, they provide 
a direct way of controlling all the singularities occurring in the Green 
functions of the theory.
Technically, divergences are generated by insertions of peculiar composite operators 
involving TF, namely the ones obtained by solving the equations of motion for NTF. 
When several external lines are non-transverse, several extra loop integrations
are generated; nevertheless we know that the corresponding
amplitudes are renormalizable as this follows from the general treatment in the 
four-component formulation.

Of course Green functions with insertions of composite operators can also be
computed by means of the operator algebra. However, using a path-integral approach,
the procedure becomes automatic and by far simpler. To realize this point the reader
should compare the situation concerning the Slavnov-Taylor identities: obviously
the physics is quite different, but the mathematical context is similar. Slavnov-
Taylor identities can be obtained using T-products, but the path-integral approach
is more convenient. That is precisely what the BI accomplish in this case. 

Since $W$ in eq.(\ref{twoco}) is gaussian in the sources $J_+,\bar \zeta_+,\zeta_+$,
it is easy to give an explicit formula for functional differentiation with respect
to those sources. Terms in $K,\bar \Theta, \Theta$, which depend on TF, can in turn be
represented in terms of functional derivatives as usual, thus leading to the following
equivalences between functional differential operators
\beeq \label{diffop}
\p_{-}^{2}\frac{\delta}{i \delta J^{+,a}}&=& J^{+,a}+ \partial_-
K^a\left[\frac{\delta}{i \delta J^{\beta}}, \frac{\delta}{i\delta \zeta_{-,m}^{b}},
\frac{\delta}{i\delta \bar{\zeta}_{-,m}^{b}}, \frac{\delta}{i\delta h_{r}^{b}},
\frac{\delta}{i\delta k_{r}^{b}}\right]= \\ \nonumber
&=&\left[J^{+,a}+
(D_{\alpha}[\frac{\delta}{i \delta J^{\beta}}]\p_{-}
\frac{\delta}{i \delta J^{\alpha}})^{a} - \frac{ig}{2}f^{abc}
\frac{\delta}{i\delta \zeta_{-,m}^{b}}\gamma^{+} 
\frac{\delta}{i\delta \bar{\zeta}_{-,m}^{c}}\right.+  \nonumber \\
&+&\left. g f^{abc}\frac{\delta}{i\delta h_{r}^{b}}\p_{-} 
\frac{\delta}{i\delta h_{r}^{c}}+ g f^{abc}\frac{\delta}{i\delta k_{r}^{b}}\p_{-} 
\frac{\delta}{i\delta k_{r}^{c}}\right],\nonumber 
\eeeq
\newpage
\beeq\label{diffop2}
\p_{-}\frac{\delta}{i \delta \bar{\zeta}_{+,m}^{a}}
&=&\frac{i}{2}\gamma^{+}\left[\zeta_{+,m}^{a}+\Theta^a_m\left[\frac{\delta}
{i \delta J^{\beta}},
\frac{\delta}{i\delta \bar{\zeta}_{-,n}^{b}}, \frac{\delta}{i\delta h_{r}^{b}},
\frac{\delta}{i\delta k_{r}^{b}}\right]\right]= \\ \nonumber
&=& \frac{i}{2}\gamma^{+}
\left[\zeta_{+,m}^{a}+i \gamma^{\alpha}
D_{\alpha}[\frac{\delta}{i \delta J^{\beta}}]\frac{\delta}{i \delta 
\bar{\zeta}_{-,m}^{a}} -g f^{abc}(\alpha_{mn}^{r} 
\frac{\delta}{i \delta h_{r}^{b}}+ \gamma_5 \beta_{mn}^{r} 
\frac{\delta}{i \delta k_{r}^{b}}) 
\frac{\delta}{i \delta \bar{\zeta}_{-,n}^{c}}\right],\\
\p_{-}\frac{\delta}{i \delta \zeta_{+,m}^{a}}
&=&\frac{i}{2}\left[\bar\zeta_{+,m}^{a}+\bar\Theta^a_m\left[\frac{\delta}
{i \delta J^{\beta}},
\frac{\delta}{i\delta {\zeta}_{-,n}^{b}}, \frac{\delta}{i\delta h_{r}^{b}},
\frac{\delta}{i\delta k_{r}^{b}}\right]\right]\gamma^+ = 
\frac{i}{2}\left[\bar\zeta_{+,m}^{a}+i(-\p_{\alpha}\frac{\delta}{i 
\delta \zeta^{a}_{-,m}}
\gamma^{\alpha}+ \right. \nonumber \\
&+& \left. g f^{bca}\frac{\delta}{i \delta \zeta_{-,m}^{b}}\gamma^{\alpha}
\frac{\delta}{i \delta J^{\alpha ,c}})
-g f^{bca}\frac{\delta}
{i \delta \zeta_{-,n}^{b}}(\alpha_{nm}^{r}\frac{\delta}{i \delta h^{c}_{r}}+
\gamma^{5}\beta^{r}_{nm}\frac{\delta}{i \delta k^{c}_{r}})\right]\gamma^{+}
\eeeq
when applied to the Green function generating functional W in eq.(\ref{twoco}).
We notice that the differentiations appearing at the right-hand side concern
only transverse sources.
Non-linearities occur with respect to TF; moreover,
if W is expressed in terms of Z, the generating functional of {\it connected}
Green functions, the identities themselves become
non-linear.  

Applications to concrete low order perturbative examples are deferred to the next section.

\section{Low order examples} 

Here we present two simple concrete examples where one can see the identities at work.

The first example we consider is the two-point Green function involving a transverse
and a longitudinal component of the vector potential $G_{\alpha +}$.
The BI allows to express it in terms of Green functions with only transverse
external lines, more precisely the identity relates an amplitude in the NTS 
(on the left side of the equality) to amplitudes in the TS (at the right side):
\beeq\label{vect}
&&\p_{-,x}^2 G_{\alpha+}^{ab}(x-y)=\p_{-,x}\p_{\beta, x}G_{\alpha\beta}^{ab}(x-y)+
gf^{bcd}\p_{-,z}G_{\alpha \beta \beta}^{acd}(x-z,z-y)_{|z=y} + \nonumber \\
&-&\frac{ig}{2} f^{dbc}\mbox{tr}\left[G^{cad}_{\lambda_+\alpha\bar\lambda_+}(x-z,
z-y)_{|z=y}\gamma^+\right]+ 6gf^{dbc}\p_{-,z}G^{cad}_{s\alpha s}(x-z,z-y)_{|z=y} ,
\eeeq
where the factor 6 in the last term takes into account the equal contributions 
of three scalars
and three pseudoscalars and the trace for the spinors also acts on the $\alpha, \beta$
matrices.

After Fourier transforming we get
\beeq\label{Four}
&&(r^+)^2\tilde G_{\alpha+}^{ab}(r)-r^+r_{\beta}\tilde G_{\alpha\beta}^{ab}(r)=
ig\mu^{4-2\omega}f^{bcd}\left[\int \frac{d^{2\omega}q}{(2\pi)^{2\omega}}q^+
\tilde G_{\alpha\beta\beta}^{acd}(r,-r-q,q)+\right.  \\
&+&\left.\int \frac{d^{2\omega}q}{2(2\pi)^{2\omega}}\mbox{tr}
\left[\tilde G^{cad}_{\lambda_+\alpha\bar\lambda_+}(r,-r+q,-q)\gamma^+\right]
+(10-2\omega) \int \frac{d^{2\omega}q}{(2\pi)^{2\omega}}q^+ 
\tilde G^{cad}_{s\alpha s}(r,-r-q,q)\right], \nonumber
\eeeq
$\mu$ being the usual dimensional regularization scale.

Figure 1 presents the one-loop contributions to 
the {\it self-energies}
for all transverse and non transverse fields.
Singular contributions to the self-energies were explicitly known up to $O(g^2)$ 
\cite{bdal}; in \cite{dal} only the one-loop gluon self-energy, including finite parts, 
has been computed. For the purpose of performing a complete check of the BI, here we 
have calculated the finite parts of the fermion self-energies as well.   
We have used the ``four-component'' 
formulation and therefore the Feynman graph topologies are the usual ones. 
The results are presented in Table 1. 

After suitable projections, we obtain the related two-point Green functions 
we summarize in Table 2. One immediately 
realizes that there is no 
divergent correction to transverse two-point Green functions.
For the transverse three-point Green functions, 
since the check is performed up to $O(g^2)$,
the tree level is enough. 

There are in principle two equivalent possibilities: to use either
the two-component formulation of the theory or the four-component one
performing the suitable projections afterwards. At this stage they are 
essentially equivalent; at higher perturbative orders the second one is
simpler thanks to its traditional Feynman graph topology.
Tree level three-point amplitudes are reported in Figure 2.

It is now fairly easy to perform the balance in eq.(\ref{Four}). First of all we
realize that the contributions from integrals of three point functions involving spinors
and scalar-pseudoscalar particles vanish. As a matter of fact we get (see Fig.2)
\beq\label{bolsp}
\int \frac {d^{2\omega}q}{(2\pi)^{2\omega}}\frac{1}{q^2 (q-r)^2}\left[q^+
(q-r)_{\alpha}+(q-r)^+q_{\alpha}-2(q-r)^+ q^+\frac{r_{\alpha}}{r^+}\right]=0
\eeq
and
\beq\label{bols}
\int \frac {d^{2\omega}q}{(2\pi)^{2\omega}}\frac{(2q
+r)^{\mu}}{q^2 (q+r)^2}=0.
\eeq
Moreover terms at ${\cal O}(g^0)$ trivially compensate.
The balance in eq.(\ref{Four}) involves therefore the three following contributions
\beq\label{a+}
(r^+)^2\tilde G_{\alpha+}^{ab}(r)=-\frac{2A r^+r_{\alpha}}{r^2}\left[\frac{1}{e}
+2+\frac{\pi^2}{3}-2 \mbox{L}_2(1-\zeta)-\frac{\zeta \log\zeta}{\zeta-1}\right],
\eeq
\beq\label{aab}
r^+r_{\beta}\tilde G_{\alpha\beta}^{ab}(r)=-\frac{2A r^+ r_{\alpha}}{r^2}\left[\frac
{\pi^2}{3}-2 \mbox{L}_2(1-\zeta)\right]
\eeq
and
\beq\label{3g}
ig\mu^{4-2\omega}f^{bcd}\int \frac{d^{2\omega}q}{(2\pi)^{2\omega}}q^+
\tilde G_{\alpha\beta\beta}^{acd}(r,-r-q,q)=-\frac{2A r^+r_{\alpha}}{r^2}\left[\frac{1}{e}
+2-\frac{\zeta \log\zeta}{\zeta-1}\right],
\eeq
where
\beq\label{def}
A\equiv \frac{i g^2 C_{Ad}}{16 \pi^2},\qquad \zeta\equiv \frac{2r^+r^-}{r^2},
\qquad \frac{1}{e}\equiv \frac{1}{2-\omega}+\log(\frac{4\pi\mu^2}{-r^2})-\gamma.
\eeq
and $\mbox{L}_2(z)$ is the Euler dilogarithm.
\vskip .5truecm

The second example concerns the two-point Green function involving a transverse
and a non-transverse component of the spinor field. The identity
in this case reads
\beeq\label{spin}
&&2\p_{-}G^{ab}_{\lambda_{-}\bar\lambda_{+}}(x-y)=-\gamma^{+}\cancel\p_{\perp}
G^{ab}_{\lambda_{+}\bar\lambda_{+}}(x-y)-g f^{acd}\gamma^{+}\gamma^{\alpha}
G^{dcb}_{\lambda_{+}\alpha\bar\lambda_{+}}(x-z,z-y)_{|z=x}+ \nonumber \\
&&-ig f^{acd}\gamma^{+}\alpha^r G^{dcb}_{\lambda_{+}r\bar\lambda_{+}}(x-z,z-y)_{|z=x}
-ig f^{acd}\gamma^{+}\beta^r\gamma_5 G^{dcb}_{\lambda_{+}5r\bar\lambda_{+}}
(x-z,z-y)_{|z=x},
\eeeq
where $G_{\lambda_{+}\alpha\bar\lambda_{+}}$ represents the three-point Green function 
involving two transverse components of the spinor and a transverse component of the vector
field, $G_{\lambda_{+}r\bar\lambda_{+}}$ and $G_{\lambda_{+}5r\bar\lambda_{+}}$ 
represent the three-point Green functions with scalars and pseudoscalars respectively.

After a Fourier transform we get
\beeq\label{FT}
&&2p^{+}\tilde G^{ab}_{\lambda_{-}\bar\lambda_{+}}(p)+\gamma^{+}\cancel p_{\perp}
\tilde G^{ab}_{\lambda_{+}\bar\lambda_{+}}(p)+ig \mu^{4-2\omega}\gamma^{+} \gamma^{\alpha}
f^{acd}\int \frac {d^{2\omega}q}{(2\pi)^{2\omega}}\tilde G^{dcb}_{\lambda_{+}\alpha\bar
\lambda_{+}}(q,p-q,p)+\nonumber \\
&&-g \mu^{4-2\omega}f^{acd}\gamma^{+}\int \frac {d^{2\omega}q}{(2\pi)^{2\omega}}
\left[\alpha^r 
\tilde G^{dcb}_{\lambda_{+}r\bar\lambda_{+}}(q,p-q,p)+
\beta^r\gamma_5 \tilde G^{dcb}_{\lambda_{+}5r\bar\lambda_{+}}(q,p-q,p)\right]=0. 
\eeeq
Taking into account that the integrals over three-point Green
functions involving scalar and pseudoscalar fields, give vanishing contributions
\beq\label{sspp}
\int\frac{d^{2\omega}q}{(2\pi)^{2\omega}}\frac{1}{q^2 (p-q)^2}
(p^+\cancel q_{\perp}-q^+\cancel p_{\perp})
=0,
\eeq
and recalling the results summarized in Tables 1 and 2, the balance in eq.(\ref{FT})
turns out to be

\beeq\label{g2-+}
&&2p^{+}\tilde G^{ab}_{\lambda_{-}\bar\lambda_{+}}(p)=g^2 C_{Ad}
\frac{p^+\cancel p_{\perp}P_-}{p^2}
\large[4 I(\omega)+8Jp^+ -8\frac{p^+}{p^2}p^{\mu}J_{\mu}+4\frac{2p^+p^-}{p^2} I(\omega)+
\\ \nonumber
&+& 8\frac{(p^+)^2}{p^2}J_+ -\frac {4}{p^2}M(2p^+p^-+p_{\perp}^2)
\large]\delta^{ab},
\eeeq
\beeq\label{g2++}
&&\gamma^+\cancel p_{\perp}\tilde G^{ab}_{\lambda_{+}\bar\lambda_{+}}(p)=-g^2 C_{Ad}\delta^{ab}\frac{p^+\cancel p_{\perp}P_-}{p^2}
\Large[4 I(\omega)+8Jp^+ -8\frac{p^+}{p^2}p^{\mu}J_{\mu}+
\\ \nonumber
&+& \frac{4}{p^2}[2(p^+)^2 J_+ - 2p_{\perp}^2 M+p_{\perp}^2 I(\omega)]\Large]
\eeeq
and
\beq\label{g3}
ig \mu^{4-2\omega}\gamma^{+} \gamma^{\alpha}
f^{acd}\int \frac {d^{2\omega}q}{(2\pi)^{2\omega}}\tilde G^{dcb}_{\lambda_{+}\alpha\bar
\lambda_{+}}(p,q,q-p)=-\frac{4g^2 C_{Ad}}{p^2}\delta^{ab}p^+\cancel p_{\perp}P_-
\large[I(\omega)-M\large],
\eeq
where $M\equiv \frac{i}{16\pi^2} \frac{\zeta\log \zeta}{\zeta -1}.$
It is fairly easy to check that $\tilde G^{ab}_{\lambda_{+}\bar\lambda_{+}}$ is
free from UV divergences.

We notice that again the presence of scalar and pseudoscalar
fields affects only two-point
functions.

\section{Concluding remarks} 

\noindent
We conclude by recalling that the light-cone quantization and renormalization of YMT
developed in the eighties, put this gauge on an equal footing with respect to other 
gauge choices. In the past it has been successfully used in phenomenological partonic 
calculations; another peculiarity concerns the
treatment of composite operators. As a matter of fact the light-cone gauge was used
in the study of evolution equations for the matrix elements of composite
quasi-partonic operators in the pioneering paper by A. Bukhvostov, G. Frolov, E. Kuraev 
and L. Lipatov \cite{bfkl}; those operators are the first example of integrable
structures in SYM $\cal{N}$=4. In ref.\cite{ba}
it has been shown that gauge invariant 
operators in QCD mix under
renormalization only among themselves  at variance with the situation occurring 
in covariant gauges.

The novelty in SYM $\cal{N}$=4 is that the coupling constant
does not run, owing to the vanishing of the $\beta$-function. 
The advantage in using the light-cone gauge in this case mainly resides in 
the direct proof 
of the finiteness of an entire sector of the theory (the TS), where supersymmetry 
is realized in 
terms of a single chiral superfield, as firstly realized by Mandelstam \cite{mand}. 
Divergences are present in the Green functions of the NTS; however they
are completely controlled by the identities we have
hitherto proposed. Indeed they only occur in connection with non transverse external 
lines: each external non transverse line entails the insertion of a composite 
operator, quadratic in the transverse fields, and thereby
a {\it single} extra loop integration over a UV-finite amplitude. 

Those divergences concern the wave-function renormalization  of NTF and, 
in the operator language, they are generated by the insertions of peculiar
composite operators, namely the ones obtained by solving the equation of motion of
``longitudinal'' operators in terms of transverse ones. The BI automatically
realize such insertions
in a path-integral formulation and might be useful, for instance, 
in higher order perturbative calculations.

To prove finiteness of the TS in the ``four component''
formulation would entail more and more cancellations in higher orders.

Since the BI are exact identities, they should survive unscathed in the presence of 
instantons; one should however keep in mind that the light cone gauge choice is mandatory,
turning the theory into a two-component formulation. We do not believe this is a 
convenient
setting for the study of instanton effects; in particular difficulties of a topological
nature might arise when inverting the operator $\p_-$. Nevertheless an insight to this
problem may be worthwhile.

Finally our procedure could be generalized to $\beta$-deformed SYM $\cal{N}$=4, like
for instance, the one considered in ref.\cite{kov}. As a matter of fact, by choosing
the light cone gauge, non transverse degrees of freedom can again be functionally integrated  
over, turning the theory into a two-component formulation. Finiteness of the Green functions
of the transverse sector however can only be proved in the planar limit.

\section*{Appendix}

We begin by providing the integrals which are needed in order to evaluate the one-loop
perturbative corrections.

\beeq
\nonumber I(\omega)&\equiv& \int \frac{d^{2\omega}q}{(2\pi)^{2\omega}}
\frac{(\mu^{2})^{2-\omega}}{q^2 (q-p)^2}
=\frac{i}{16 {\pi}^{2}}(-\frac{p^2}{4 \pi \mu^{2}})^{\omega -2}
\frac{\Gamma(2-\omega)\Gamma^{2}(\omega -1)}{\Gamma (2\omega -2)}, \\ \nonumber
I_{\mu}& \equiv &\int \frac{d^{2\omega}q}{(2\pi)^{2\omega}}
\frac{q_{\mu}(\mu^{2})^{2-\omega}}{q^2 (q-p)^2}=\frac{p_{\mu}}{2}I(\omega),\\
\nonumber
I_{\mu \nu}& \equiv &\int \frac{d^{2\omega}q}{(2\pi)^{2\omega}}
\frac{q_{\mu}q_{\nu}(\mu^{2})^{2-\omega}}{q^2 (q-p)^2}=I(\omega)
\frac{1}{4(2\omega-1)}\left[2\omega p_{\mu}p_{\nu}-g_{\mu \nu}p^2\right],\\ 
\nonumber
J& \equiv & \int \frac{d^4q}{(2\pi)^4}
\frac{1}{q^2 q^+ (q-p)^2}=\frac{i}{16 \pi^2 p^+}[\frac{\pi^2}{6}-\mbox{L}_2
(1-\zeta)],\\ \nonumber
J_{\alpha}& \equiv & \int \frac{d^4q}{(2\pi)^4}
\frac{q_{\alpha}}{q^2 q^+ (q-p)^2}=\frac{i p_{\alpha}}{4 (2\pi)^2}\frac{1}{p^+}
\frac{\zeta\log \zeta}{\zeta -1},\\ \nonumber 
J_{+}& \equiv & \int \frac{d^4q}{(2\pi)^4}
\frac{q_{+}}{q^2 q^+ (q-p)^2}=\frac{ip^2}{32 \pi^2 (p^+)^2}\large[\frac {\pi^2}{6}
-\mbox{L}_2(1-\zeta)+\zeta \log \zeta-\zeta \large],\\ \nonumber
J_{-}& \equiv & \int \frac{d^{2\omega}q}{(2\pi)^{2\omega}}\frac{\mu^{4-2\omega}
q_{-}}{q^2 q^+ (q-p)^2}=I(\omega).\nonumber
\eeeq

Now we proceed by presenting in Fig.1 all the one-loop contributions to the 
{\it self-energies} for transverse and non transverse fields. 
Those self-energies are computed using the ``four-component'' 
formulation; 
therefore the Feynman graph topologies are the usual ones. The results are
reported in Table 1 together with the relevant definitions. 

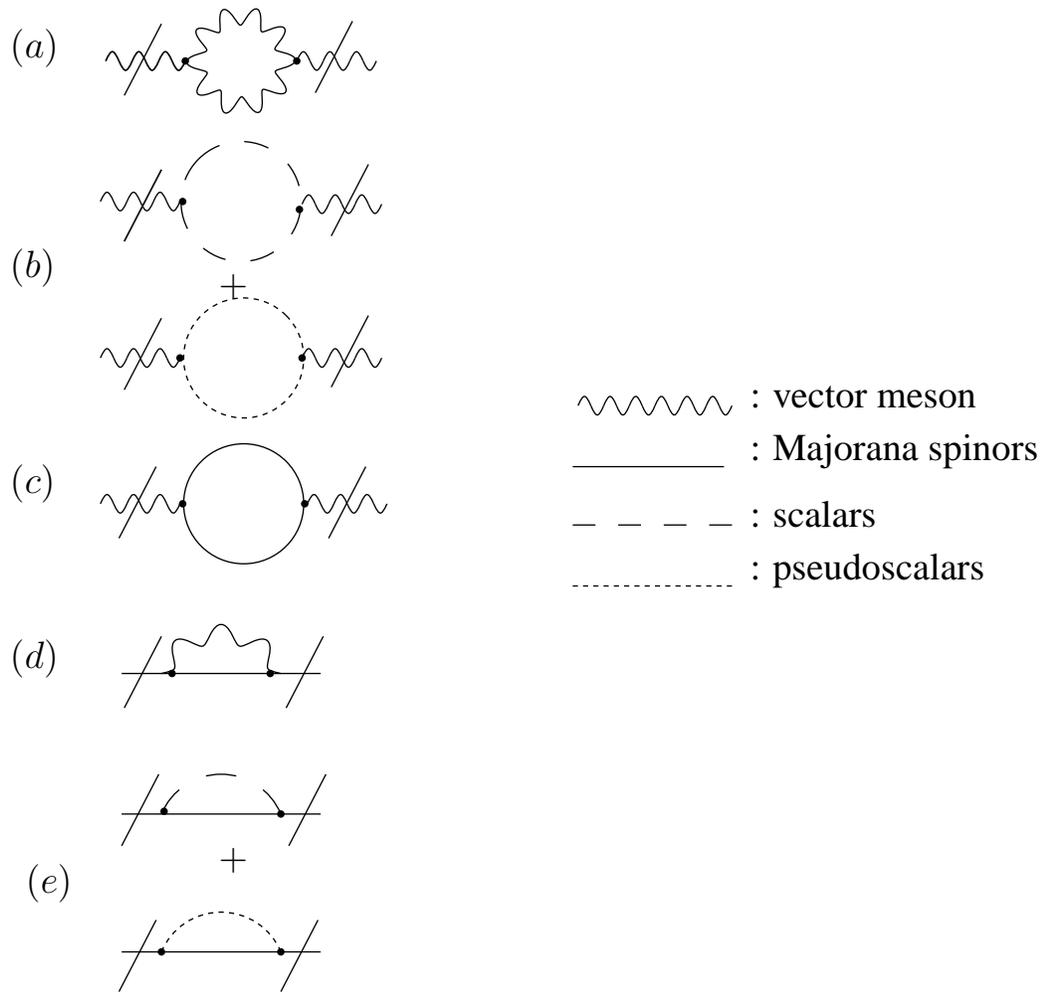
\begin{figure}[ht]
\begin{center}
  \begin{picture}(310,371) (10,-19)
    \SetWidth{0.5}
    \PhotonArc(97,331)(17,-152,208){3.5}{10}
    \Photon(46,331)(76,331){3.5}{3}
    \Photon(118,331)(148,331){3.5}{3}
    \Vertex(76,331){1.41}
    \Vertex(118,331){1.41}
    \Photon(46,331)(76,331){3.5}{3}
    \Photon(44,278)(74,278){3.5}{3}
    \Vertex(75,278){1.41}
    \DashCArc(97,278)(22.63,135,495){10}
    \Photon(44,219)(74,219){3.5}{3}
    \DashCArc(98,219)(22.63,45,405){2}
    \Vertex(74,219){1.41}
    \CArc(98,164)(22.63,45,405)
    \Photon(44,164)(74,164){3.5}{3}
    \Vertex(75,164){1.41}
    \Photon(122,164)(152,164){3.5}{3}
    \Vertex(121,164){1.41}
    \Photon(120,219)(150,219){3.5}{3}
    \Vertex(120,219){1.41}
    \Vertex(119,275){1.41}
    \Photon(120,277)(150,277){3.5}{3}
    \Line(52,100)(127,100)
    \PhotonArc(89.5,90.63)(24.37,22.6,157.4){-3.5}{3.5}
    \Line(52,47)(127,47)
    \DashCArc(89.5,37.63)(24.37,22.6,157.4){10}
    \Line(52,-5)(127,-5)
    \DashCArc(89.5,-14.37)(24.37,22.6,157.4){2}
    \Vertex(71,100){1.41}
    \Vertex(108,100){1.41}
    \Vertex(68,48){1.41}
    \Vertex(112,47){1.41}
    \Vertex(67,-5){1.41}
    \Vertex(112,-5){1.41}
    \Text(89,25)[lb]{\Large{$+$}}
    \Text(89,242)[lb]{\Large{$+$}}
    \Text(10,330)[lb]{\Large{$(a)$}}
    \Text(10,247)[lb]{\Large{$(b)$}}
    \Text(10,165)[lb]{\Large{$(c)$}}
    \Text(10,99)[lb]{\Large{$(d)$}}
    \Text(16,14)[lb]{\Large{$(e)$}}
    \Line(53,318)(67,345)
    \Line(125,319)(139,346)
    \Line(53,263)(67,290)
    \Line(53,263)(67,290)
    \Line(131,265)(145,292)
    \Line(53,207)(67,234)
    \Line(131,208)(145,235)
    \Line(51,151)(65,178)
    \Line(130,151)(144,178)
    \Line(53,87)(67,114)
    \Line(114,87)(128,114)
    \Line(52,35)(66,62)
    \Line(115,35)(129,62)
    \Line(51,-20)(65,7)
    \Line(112,-20)(126,7)
    \Photon(224,201)(282,201){3.5}{6}
    \Line(222,178)(279,178)
    \DashLine(222,156)(282,156){10}
    \DashLine(222,133)(282,133){2}
    \Text(290,201)[lb]{\Large{$\mbox{: vector meson}$}}
    \Text(290,156)[lb]{\Large{$\mbox{: scalars}$}}
    \Text(290,133)[lb]{\Large{$\mbox{: pseudoscalars}$}}
    \Text(290,178)[lb]{\Large{$\mbox{: Majorana spinors}$}}
  \end{picture}
\end{center}

\caption{One loop self-energy diagrams\label{bolle}}
\end{figure}
\begin{table}[!h]
\begin{center}
\begin{tabular}[!h]{c c}\hline \hline 
$(a)$ & $\{ (\frac{11}{3 e}+\frac{67}{9}) S^{\mu \nu}_{(1)}+
2\left[-2(\frac{\pi^2}{6}-\mbox{L}_2(1-\zeta))(S^{\mu \nu}_{(1)}+
S^{\mu \nu}_{(2)})\right.$\\
$$ &$\left.+2 \frac{\zeta \log \zeta}{\zeta -1}S^{\mu \nu}_{(2)}
+(\frac{1}{e}+2-\frac{\zeta \log \zeta}{\zeta -1})S^{\mu \nu}_{(3)}\right]
\}A$\\
$(b)$ & $-\left[\frac{1}{e}+3\right] S^{\mu \nu}_{(1)} A$\\
$(c)$ & $-\frac{8}{3}\left[\frac{1}{e}+\frac{5}{3}\right] 
S^{\mu \nu}_{(1)} A$\\
$(a)+(b)+(c)$&$2\left[-2(\frac{\pi^2}{6}-\mbox{L}_2(1-\zeta))(S^{\mu \nu}_{(1)}+
S^{\mu \nu}_{(2)})+2 \frac{\zeta \log \zeta}{\zeta -1}S^{\mu \nu}_{(2)}
+(\frac{1}{e}+2-\frac{\zeta \log \zeta}{\zeta -1})S^{\mu \nu}_{(3)}\right]A$ \\ \hline
$(d)$ & $\delta^{m_1 m_2} C_{Ad}g^2 \large[(\omega-3)\cancel p I(\omega)+2
\cancel n p^2 J -2 \cancel n p^{\mu}J_{\mu}+2 p^+ \cancel J \large]$\\
$(e)$ & $(5-\omega) \delta^{m_1 m_2}C_{Ad}g^2 \cancel{p} I(\omega)$\\
$(d)+(e)$ & $\delta^{m_1 m_2} C_{Ad}g^2 \large[2\cancel p I(\omega)+2
\cancel n p^2 J -2 \cancel n p^{\mu}J_{\mu}+2 p^+ \cancel J \large]$
\\ \\
\hline \hline
\multicolumn{2}{c}{$A=i g^2 C_{Ad} \delta^{a_1 a_2} / 16 \pi^{2} $ }\\
\multicolumn{2}{c}{$C_{Ad} \delta^{a_1 a_2}= f^{a_1 b c} f^{a_2 b c}$} \\
\multicolumn{2}{c}{$S_{(1)}^{\mu \nu}=g^{\mu \nu} p^2 -p^{\mu} p^{\nu}$} \\
\multicolumn{2}{c}{$S_{(2)}^{\mu \nu}=p^{\mu}p^{\nu}-(n^{\mu}p^{\nu}+
n^{\nu}p^{\mu}) p^2/n \cdot p +n^{\mu}n^{\nu}(p^2)^2/(n \cdot p)^2$}\\
\multicolumn{2}{c}{$S_{(3)}^{\mu \nu}=[ 2 n^{\mu} n^{\nu} p^2 \hat{n}\cdot p /n \cdot p +n \cdot p (\hat{n}^{\mu} p^{\nu}+\hat{n}^{\nu}p^{\mu})-
\hat{n}\cdot p (n^{\mu} p^{\nu}+n^{\nu} p^{\mu}) -p^2 (n^{\mu} \hat{n}^{\nu}+
n^{\nu} \hat{n}^{\mu})]/n \cdot \hat{n}$  }\\
\multicolumn{2}{c}{$\cancel{B}=(\cancel{\hat{n}}\, n \cdot p-\cancel{n}\,
\hat{n}\cdot p)/n \cdot \hat{n}$}\\ \hline \hline

\end{tabular}
\caption{One loop self-energies. \label{partidivergenti}}
\end{center}
\end{table}

Table 2 presents the corresponding Green functions at $O(g^0)$ and at 
$O(g^2)$. The three-point Green 
functions at tree level for the transverse fields, which are  required
to match at $O(g^2)$ the BI, appear in Fig. 2.

\begin{table}[!h]
\begin{center}
\begin{tabular}[!h]{c c}\hline \hline

$G^{(0)}_{\lambda_{+} \bar{\lambda}_{+}}=P_+ S^{(0)} P_-=
\frac{i p^+ \gamma^{-}}{p^2}$\\

$G^{(0)}_{\lambda_{+} \bar{\lambda}_{-}}=P_+ S^{(0)} P_+=
\frac{i \cancel{p}_{\perp}}{p^2} P_+$\\

$G^{(0)}_{\lambda_{-} \bar{\lambda}_{+}}=P_- S^{(0)} P_-=
\frac{i \cancel{p}_{\perp}}{p^2} P_-$\\

$G^{(0)}_{\lambda_{-} \bar{\lambda}_{-}}=P_- S^{(0)} P_+=
\frac{i p^- \gamma^{+}}{p^2}$\\


$G^{(2)}_{\alpha \beta}=\frac{4 A}{p^2}g_{\alpha \beta}\left[\frac{\pi^2}{6}
-\mbox{L}_2(1-\zeta)\right]$\\

$G^{(2)}_{\alpha +}=-\frac{2A}{p^2}\frac{p_{\alpha}}{p^+}\left[
\frac{1}{2-\omega}+\log (\frac{4 \pi \mu^2}{-p^2})-\gamma +2
+2(\frac{\pi^2}{6}
-\mbox{L}_2(1-\zeta))-\frac{\zeta\log \zeta}{\zeta -1}\right]$\\

$G^{(2)}_{\lambda_{+} \bar{\lambda}_{+}}=g^2 C_{Ad}\large[\frac{4(p^+)^2}{p^4} p^{\alpha}J_{\alpha}
-\frac{4(p^+)^2}{p^2}J+\frac{4p^+ p_{\perp}^2}{p^4}M\large]\gamma^-$\\


$G^{(2)}_{\lambda_{-} \bar{\lambda}_{+}}=-g^2 C_{Ad}\large[\frac{2}{p^2}I(\omega)+
\frac{4p^+}{p^2}J-\frac{4p^+}{p^4}p^{\alpha}J_{\alpha}-\frac{4p^+p^-+2p_{\perp}^2}{p^4}
M\large]\cancel p_{\perp}P_-
$\\


\hline
\hline

\end{tabular}
\caption{0- and 1-loop two-point Green functions \label{greenf}}
\end{center}
\end{table}

\begin{figure}[ht]
\begin{center}

  \begin{picture}(152,102) (35,-7)
    \SetWidth{0.5}
    \Photon(116,42)(158,94){6}{4}
    \ArrowLine(154,64)(140,48)
    \Text(69,58)[lb]{\Large{$\alpha\,,\,a$}}
    \Text(55,29)[lb]{\Large{$p$}}
    \Text(105,71)[lb]{\Large{$\gamma\,,\,c$}}
    \Text(157,72)[lb]{\Large{$r$}}
    \Text(157,13)[lb]{\Large{$\beta\,,\,b$}}
    \Text(133,-5)[lb]{\Large{$q$}}
    \Photon(116,42)(35,44){6}{4}
    \Photon(115,42)(164,-8){6}{4}
    \ArrowLine(64,28)(90,28)
    \ArrowLine(133,8)(118,24)
  \end{picture}

$\tilde{G}^{(1)}_{\alpha \beta \gamma}=\frac{i g f^{abc}}{p^2 q^2 r^2}
\left[ \delta_{\alpha \beta}(p-q)_{\gamma}+
\delta_{\beta \gamma}(q-r)_{\alpha}+
\delta_{\gamma \alpha}(r-p)_{\beta}+\right.$\\
$\left. -\delta_{\alpha \beta}\frac{r_{\gamma}}{r^+} (p^+ -q^+)-
\delta_{\beta \gamma}\frac{p_{\alpha}}{p^+} (q^+ -r^+)-
\delta_{\gamma \alpha}\frac{q_{\beta}}{q^+} (r^+ -p^+)
\right]$

\vskip .5truecm

  \begin{picture}(152,102) (35,-7)
    \SetWidth{0.5}
    \Photon(116,42)(158,94){6}{4}
    \ArrowLine(154,64)(140,48)
    \Text(69,58)[lb]{\Large{$a$}}
    \Text(55,29)[lb]{\Large{$p$}}
    \Text(105,71)[lb]{\Large{$\alpha\,,\,c$}}
    \Text(157,72)[lb]{\Large{$r$}}
    \Text(157,13)[lb]{\Large{$b$}}
    \Text(133,-5)[lb]{\Large{$q$}}
    \Line(116,42)(35,44)
    \Line(115,42)(164,-8)
    \ArrowLine(64,28)(90,28)
    \ArrowLine(118,24)(133,8)
  \end{picture}

$\tilde{G}^{(1)}_{\lambda_{+}\alpha \bar{\lambda}_{+}}=\frac{i g f^{acb}}
{p^2 q^2 r^2}\left[
q^+ \cancel{p}_{\perp}\gamma_{\alpha}+
p^+ \gamma_{\alpha} \cancel{q}_{\perp} 
-2 \frac{p^+ q^+}{r^+}r_{\alpha}\right]\gamma^{-}$

\vskip .5truecm

  \begin{picture}(152,102) (35,-7)
    \SetWidth{0.5}
    \Photon(116,42)(158,94){6}{4}
    \ArrowLine(154,64)(140,48)
    \Text(69,58)[lb]{\Large{$a$}}
    \Text(55,29)[lb]{\Large{$p$}}
    \Text(105,71)[lb]{\Large{$\alpha\,,\,c$}}
    \Text(157,72)[lb]{\Large{$r$}}
    \Text(157,13)[lb]{\Large{$b$}}
    \Text(133,-5)[lb]{\Large{$q$}}
    \DashLine(116,42)(35,44){10}
    \DashLine(115,42)(164,-8){10}
    \ArrowLine(64,28)(90,28)
    \ArrowLine(133,8)(118,24)
  \end{picture}

$\tilde{G}^{(1)}_{s\alpha s}=\frac{i g f^{acb}}{p^2 q^2 r^2}
\left[\frac{r_{\alpha}}{r^+}(p^+ -q^+)
-(p_{\alpha}-q_{\alpha})\right]$

\vskip .5truecm

  \begin{picture}(152,102) (35,-7)
    \SetWidth{0.5}
    \DashLine(116,42)(158,94){10}
    \ArrowLine(154,64)(140,48)
    \Text(69,58)[lb]{\Large{$a$}}
    \Text(55,29)[lb]{\Large{$p$}}
    \Text(105,71)[lb]{\Large{$c$}}
    \Text(157,72)[lb]{\Large{$r$}}
    \Text(157,13)[lb]{\Large{$b$}}
    \Text(133,-5)[lb]{\Large{$q$}}
    \Line(116,42)(35,44)
    \Line(115,42)(164,-8)
    \ArrowLine(64,28)(90,28)
    \ArrowLine(118,24)(133,8)
  \end{picture}

$\tilde{G}^{(1)}_{ \lambda_{+}s \bar{\lambda}_{+}}=g f^{acb} \alpha^{r}_{mn}
\frac{1}{p^2 q^2 r^2} (p^+ \cancel{q}_{\perp}-q^+ \cancel{p}_{\perp})
\gamma^{-}$

\end{center}
\caption{Tree-level Green functions\label{greenfunctions}}
\end{figure}

\pagestyle{plain}

\end{document}